\documentclass[aps,prl,twocolumn,floatfix,showpacs,groupaddress,longbibliography,nofootinbib]{revtex4-2}

\usepackage{amssymb}
\usepackage{amsmath}
\usepackage{bm}
\usepackage{graphicx}
\usepackage{hyperref}
\usepackage{xcolor}
\usepackage{soul}

\usepackage[utf8]{inputenc}

\begin{document}

\title{Active Nematic Ratchet in Asymmetric Obstacle Arrays}

\author{Cody D. Schimming}
\email[]{cschim@lanl.gov}
\affiliation{Theoretical Division and Center for Nonlinear Studies, Los Alamos National Laboratory, Los Alamos, New Mexico, 87545, USA}

\author{C. J. O. Reichhardt}
\affiliation{Theoretical Division and Center for Nonlinear Studies, Los Alamos National Laboratory, Los Alamos, New Mexico, 87545, USA}

\author{C. Reichhardt}
\affiliation{Theoretical Division and Center for Nonlinear Studies, Los Alamos National Laboratory, Los Alamos, New Mexico, 87545, USA}

\begin{abstract}

We numerically investigate the effect of a periodic array of asymmetric obstacles in a two-dimensional active nematic. We find that activity in conjunction with the asymmetry leads to a ratchet effect or unidirectional flow of the fluid along the asymmetry direction. The directional flow is still present even in the active turbulent phase when the gap between obstacles is sufficiently small. We demonstrate that the dynamics of the topological defects transition from flow-mirroring to smectic-like as the gap between obstacles is made smaller, and explain this transition in terms of the pinning of negative winding number defects between obstacles. This also leads to a non-monotonic ratchet effect magnitude as a function of obstacle size, so that there is an optimal obstacle size for ratcheting at fixed activity.
\end{abstract}

\maketitle

Active nematics are anisotropic fluids that exhibit local orientational order and generate macroscopic flows from microscopic forces \cite{marchetti13,doo18}. In large, unconfined systems these flows are typically chaotic, leading to a phase dubbed ``active turbulence'' \cite{Sanchez12,doo17,alert20,carenza20}. Additionally, orientational order in the nematic allows the existence of topological defects, which may spontaneously nucleate in the active turbulence phase and act as sources for the flow \cite{giomi14,DeCamp15,Shankar19,angheluta21,ronning22}. In addition to the inherent interest in
chaotic flows and defect dynamics in active nematic turbulence,
much recent activity has focused on
controlling the flows
for potential technological and biological applications
such as microfluidic devices, wound healing, and morphogenesis \cite{woodhouse17,guillamat22,hoffman22}.
Proposed flow control methods include modifying the boundary geometry, employing spatially varying activity, applying external fields, and altering substrate properties \cite{guillamat16,guillamat16b,shendruk17,opathalage19,thijssen20,thijssen21,Rzhang22,zarei23}. 
There has also been experimental work on the interaction of
active nematics with fabricated obstacle arrays \cite{figueroa22,velez24}, where
defect pinning was observed.

A ratchet effect can be used to control flows in
systems coupled to an asymmetric substrate under external ac driving or
flashing of the substrate
\cite{Magnasco93,Astumian94,Bartussek94,Reimann02}.
Ratchet effects have been demonstrated for colloidal particles
\cite{Rousselet94, Arzola11}
and superconducting vortices \cite{Lee99,deSouzaSilva06a},
where ac driving results in a net unidirectional flow of particles.
In active matter systems coupled to
asymmetric substrates, ratchet effects
can arise without external driving
due to the activity \cite{Bechinger16,Reichhardt17a}.
Particle-based active matter ratchets have been studied
for biological systems such as swimming bacteria \cite{Galajda07}
as well as active colloids \cite{Nikola16,Reichhardt17a}.
An open question is whether ratchet effects also occur for
active nematics coupled to an asymmetric substrate,
and if so, how the fluid flow and topological defects would be modified.

Here, we numerically study a
two-dimensional active nematic interacting with a
periodic array of asymmetric obstacles of triangular shape.
Topological defects, which are known to generate flows \cite{doo18,ronning22},
spontaneously
appear in the system out of geometrical necessity due to the shape of the
obstacles.
We show that when the gap between the asymmetric obstacles is
sufficiently small, an active nematic ratchet effect occurs
in the form of unidirectional flow along the asymmetry axis,
something that does not occur for an array of symmetric obstacles
\cite{schimming24}.
Ratcheting effects have been observed for rotational flows in active nematics
interacting with asymmetric inclusions and boundaries \cite{wu17,ray23}, but, to our knowledge, this is the first realization of a translational active nematic ratchet. We demonstrate that the ratchet effect is robust across a wide range of obstacle gap sizes and activity levels.
By tuning the gap size, a transition in the defect dynamics occurs, and the
flow speed is optimized at the transition point.

We model a two-dimensional active nematic using a well-established nemato-hydrodynamics model in terms of the tensor order parameter $\mathbf{Q} = S\left[\mathbf{n}\otimes\mathbf{n} - (1/2)\mathbf{I}\right]$, where $S$ is the local degree of orientational order and the director, $\mathbf{n}$, gives the local orientation of the nematic \cite{marenduzzo07,doo18,SuppNote24}. We measure lengths and times in units of the nematic correlation length $\xi$ and the nematic relaxation time $\tau$, respectively, so the dimensionless evolution equation for $\mathbf{Q}$ is given by
\begin{equation} \label{eqn:QEvo}
    \frac{\partial \mathbf{Q}}{\partial t} + \left(\mathbf{v} \cdot \nabla\right)\mathbf{Q} - \mathbf{S} = -\frac{\delta F}{\delta \mathbf{Q}}
\end{equation}
where $\mathbf{v}$ is the fluid velocity, $\mathbf{S}$ is a generalized tensor advection, and $F$ is a Landau-de Gennes free energy with a single elastic constant \cite{deGennes75,beris94,SuppNote24}.
We work in a free energy regime where
the passive nematic is in the nematic phase and the equilibrium defect diameter is unity. The fluid velocity is generated from active stresses given by inhomogeneities in the nematic, and is computed from the Stokes equation:
\begin{equation} \label{eqn:Stokes}
    \nabla^2 \mathbf{v} = \nabla p + \alpha \nabla \cdot \mathbf{Q}
\end{equation}
where $p$ is the fluid pressure and $\alpha$ is the strength of active forces, called the activity. We also assume the fluid is incompressible and enforce the constraint $\nabla \cdot \mathbf{v} = 0$. 

\begin{figure}
\centering
   \includegraphics[width = \columnwidth]{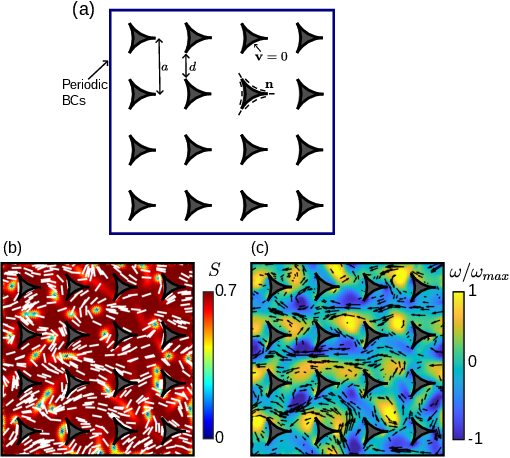}
   \caption{(a) Schematic of the computational domain with a periodic array of triangular obstacles. (b) Time snapshot of the nematic configuration. The color represents the scalar order parameter $S$ while the white lines represent the director $\mathbf{n}$. (c) Time snapshot of the vorticity and velocity. The color represents the vorticity while the black lines represent the flow field.}
    \label{fig:TriObstacleDomain}
\end{figure}

We discretize Eqs. \eqref{eqn:QEvo} and \eqref{eqn:Stokes} in space and time and solve them using the MATLAB/C++ package FELICITY \cite{walker18,SuppNote24}. We simulate on domains with a square lattice of concave triangular obstacles that break the symmetry along the $x$-axis, shown in Fig. \ref{fig:TriObstacleDomain}(a). We use strong planar anchoring of the director on the obstacles, as well as a no slip condition for the fluid velocity. On the outer boundaries, we employ periodic boundary conditions. The distance between obstacle centers is fixed at $a = 14$ while we vary the size of the obstacles so that the shortest gap between them, $d$, changes. Due to the strong planar anchoring, the concave triangles each carry a topological charge (winding number) of $-1/2$. The total topological charge of the system must be $0$ due to the periodic boundary conditions, so a defect of charge $+1/2$ must nucleate in the bulk nematic for each obstacle. We show in Fig. \ref{fig:TriObstacleDomain}(b) a time snapshot of the scalar order parameter $S$ and director field $\mathbf{n}$ for a system with $\alpha = 1$ and $d = 4$, while Fig. \ref{fig:TriObstacleDomain}(c) shows the corresponding time snapshot of the velocity and vorticity field.

\begin{figure}
\centering
    \includegraphics[width = \columnwidth]{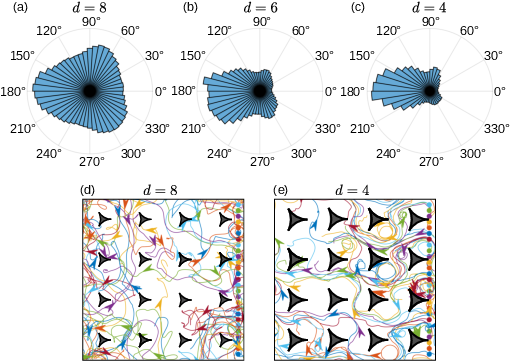}
\caption{(a,b,c) Distribution of velocity directions $p(\theta_v)$ for
varied gap width $d$ at activity $\alpha=1$.
(a) $d=8$, (b) $d=6$, (c) $d=4$.  
(c,d) Trajectories of 27 virtual particles over the course of a simulation with
varied $d$ at $\alpha = 1$. The dots on the right side indicate the starting positions of the particles.
(c) $d=8$. (d) $d=4$.
}
    \label{fig:VelocityDistributions}
\end{figure}

At $\alpha = 0$, there are no flows in the system and topological defects are pinned to the obstacles. For $\alpha > 0$, topological defects unpin and move while additional defects continuously nucleate and annihilate. For all obstacle gap sizes, we find that the average number of defects in the system and the average magnitude of the flow velocity increases linearly with the activity (Fig.~S1). These are traditional measures of active turbulence in active nematic systems \cite{DeCamp15,lemma19,opathalage19}, indicating that the system is in active turbulence for $\alpha > 0$. This differs significantly from our recent study on active nematics in periodic arrays of symmetric obstacles, where multiple phase transitions occurred when varying $\alpha$ \cite{schimming24}. 

Although the flow measurements in our system are consistent with active
turbulence, the detailed nature of the flow differs from
traditional active turbulent states, where
the flow directions are distributed randomly.
We find that the flows through our asymmetric obstacles are distributed anisotropically,
as illustrated in Fig.~\ref{fig:VelocityDistributions}(a-c) where we plot
the distribution of fluid flow directions $p(\theta_v)$ in systems with
$\alpha=1$.
When the obstacle gap is large, as shown
in Fig.~\ref{fig:VelocityDistributions}(a) for $d=8$, 
the flow directions match the three-fold symmetry
directions of the obstacle surfaces, suggesting that the obstacles are
merely locally modulating the flow.
As $d$ decreases, however,
$p(\theta_v)$ becomes strongly peaked along $\theta_v = \pi$, as shown
in Figs.~\ref{fig:VelocityDistributions}(b,c) for $d=6$ and $d=4$, respectively.
This indicates the emergence of a directional or rectified flow in the absence
of an external drive.
To visualize the rectification of the flow,
in Figs.~\ref{fig:VelocityDistributions}(d,e) we plot the trajectories of $27$ virtual tracer particles that are initially placed near the right outer boundary of the domain and are advected by the flow over the course of a simulation. In Fig.~\ref{fig:VelocityDistributions}(d), for the wide gap case of $d=8$ where
strong rectification is not present,
the tracer particles generally remain close to their starting points and
have no coordinated motion.
In contrast, for $d=4$ in Fig. \ref{fig:VelocityDistributions}(e), the tracer particles tend to travel towards the left side of the domain, as indicated by the
appearance of a gradient in the density of the trajectories and regions of
aligned flow.

\begin{figure}
\centering
    \includegraphics[width = \columnwidth]{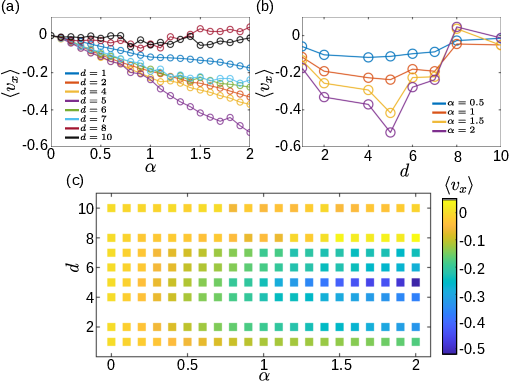}
\caption{(a) Average $x$ velocity of the flow field $\langle v_x\rangle$ vs activity $\alpha$ for $d=1$, 2, 4, 5, 6, 7, 8, and 10.
  (b) $\langle v_x \rangle$ vs $d$ for representative $\alpha$
  values of $\alpha=0.5$, 1, 1.5, and 2.0.
  (c) Heatmap of $\langle v_x \rangle$ plotted as a function of $\alpha$ vs $d$.}
    \label{fig:UxAverages}
\end{figure}

To further quantify the unidirectional flow we measure the space and time averaged $x$-component of the flow velocity $\langle v_x \rangle$. Figure \ref{fig:UxAverages}(a) shows $\langle v_x \rangle$ versus activity $\alpha$ for a range of $d$ values. When $d \geq 8$, $\langle v_x \rangle \sim 0$ for all $\alpha$, indicating that there is no net flow in the $x$-direction. For $d \leq 7$, we find that the magnitude of $\langle v_x \rangle$ increases linearly with increasing $\alpha$. Since the sign of $\langle v_x\rangle$ is negative, this indicates that there is a net flow to the left that becomes greater as the activity increases. Interestingly, we find the magnitude of $\langle v_x \rangle$ varies non-monotonically with $d$ at fixed $\alpha$, as illustrated in Fig.~\ref{fig:UxAverages}(b). To identify the overall greatest magnitude of the average flow,
we plot a heatmap of $\langle v_x\rangle$ as a function of $\alpha$ versus
$d$ in Fig.~\ref{fig:UxAverages}(c),
and find that the maximum ratchet effect occurs
for the largest simulated value of $\alpha$,
$\alpha=2$, 
at $d=5$.

There are previous studies that have shown that breaking rotational symmetry in a circular or annular domain or along a circular inclusion may induce unidirectional azimuthal flows \cite{wu17,ray23}. Further, it has been shown in channel geometries for small values of the activity that unidirectional flows may occur \cite{doo16b}; however, the flow direction is a spontaneously broken symmetry, and may be in either direction along the channel. Additionally, at higher activities, the emergence of either vortex lattices or active turbulence destroy the unidirectional flow \cite{doo16b}. For the triangular obstacles studied here, the direction of the active nematic ratchet flow is set by the asymmetry of the obstacle. To our knowledge, this is the first observation of translational active nematic ratcheting behavior. Further, below a critical gap size, the ratchet effect is robust to activity level and obstacle gap size, indicating that it would not be necessary to extensively tune the system parameters to obtain ratcheting motion for microfluidic applications.

\begin{figure}
\centering
    \includegraphics[width = \columnwidth]{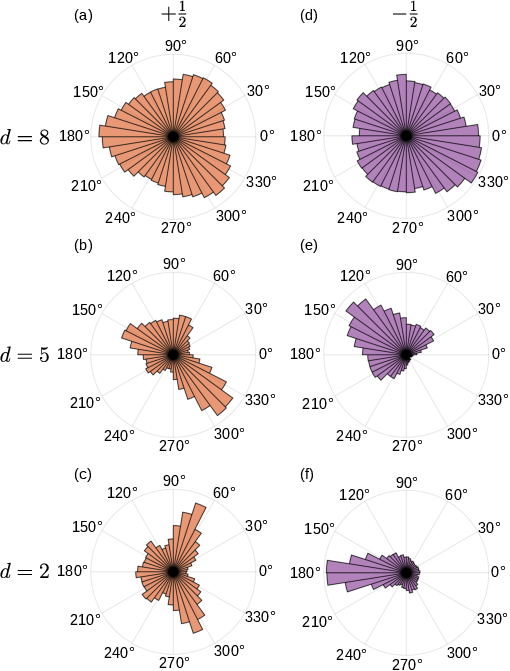}
\caption{Defect velocity direction distributions (a-c) $p(\theta_{+})$ for $+1/2$ defects and (d-f) $p(\theta_{-})$ for $-1/2$ defects at $\alpha=1.5$ for (a,d) $d = 8$, (b,e) $d = 5$, and (c,f) $d = 2$.}
    \label{fig:DefectDistributions}
\end{figure}

We also find that the defect dynamics change depending on
the obstacle gap size.
For large $d$, the plot of the distribution $p(\theta_{+})$ of
the velocities of $+1/2$ winding defects
in Fig.~\ref{fig:DefectDistributions}(a) for $d=8$ and $\alpha=1.5$ indicates
that the $+1/2$ winding defects move in the same direction as the flow.
At the same time, Fig.~\ref{fig:DefectDistributions}(d) indicates that
the $-1/2$ winding defect velocity distribution, $p(\theta_{-})$, is
much more isotropic.
As the gap size decreases, the ratchet effect emerges and the
net flow velocity is primarily along
the $-x$ direction, but for $d=5$ 
the $+1/2$ defect velocities break the up-down symmetry of the domain, as shown by the plot of $p(\theta_{+})$ in Fig. \ref{fig:DefectDistributions}(b).
In this intermediate regime, the $-1/2$ defects also break up-down symmetry and tend to move in the direction opposite to the primary flow direction of the
$+1/2$ defects, as illustrated
by the plot of $p(\theta_{-})$ in Fig.~\ref{fig:DefectDistributions}(e). In the limit of small $d$,
shown in the plots of $p(\theta_{+})$ and
$p(\theta_{-})$ in Figs.~\ref{fig:DefectDistributions}(c) and (f)
at $d=2$, the $+1/2$ defects tend to move either up or down with equal frequency, restoring the up-down symmetry of the domain, while the $-1/2$ defects primarily move to the left, in the direction of the ratcheting flow. We note that in this regime, the $+1/2$ defects tend to move transverse to the fluid flow direction.

\begin{figure}
\centering
    \includegraphics[width = \columnwidth]{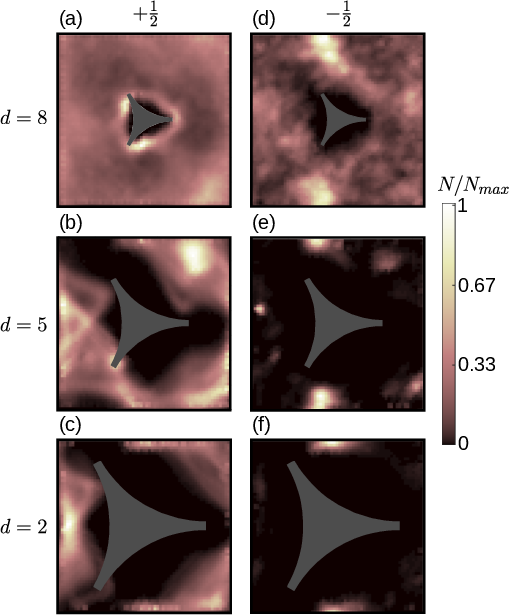}
    \caption{Distribution of defects around obstacles. The distribution around each obstacle in a simulation is computed and then all are added together to form the plots. (a--c) Distribution of $+1/2$ defects and (d--f) distribution of $-1/2$ defects for simulations with (a,d) $d = 8$, (b,e) $d = 5$, (c,f) $d = 2$ and $\alpha = 1.5$.}
    \label{fig:DefectDensities}
\end{figure}

To better understand the defect dynamics, in Fig.~\ref{fig:DefectDensities}(a--c) we construct defect density plots $N/N_{\rm max}$ as a function of position relative to the obstacle for the $+1/2$ defects at different values of $d$, and show the corresponding $N/N_{\rm max}$ plots for the $-1/2$ defects in Fig.~\ref{fig:DefectDensities}(d--f).
Here, we first accumulate $N$, the local defect density around each obstacle, on a grid surrounding the obstacle during the entire simulation. We then sum this quantity over all obstacles and normalize it by the maximum value $N_{\rm max}$ on the grid \cite{SuppNote24}.
For $d = 8$ in Figs.~\ref{fig:DefectDensities}(a,d), both the positive and negative defects are distributed throughout the interdefect region, with peak values of $N/N_{\rm max}$ appearing close to the obstacle for the $+1/2$ defects.
Thus, for large gap sizes, defects of both signs are freely moving in the domain, but the $+1/2$ defects can become briefly pinned by the obstacles.
For $d = 5$  in Figs.~\ref{fig:DefectDensities}(b,e), $N/N_{\rm max}$ for the $+1/2$ defects breaks the up-down symmetry of the obstacle and is largest along a line in interstitial space connecting the left and right sides of the obstacle, indicating that the defects are flowing horizontally. Further, there are no longer strong peaks in $N/N_{\rm max}$ near the obstacle, indicating that the $+1/2$ defects no longer become pinned (see Fig.~S2 for the distribution of radial distances of defects). At the same time, the distribution of negative defects becomes highly concentrated in the region between the upper and lower sides of the obstacles, indicating defect localization in this area.
At $d = 2$ in Figs.~\ref{fig:DefectDistributions}(c,f), $N/N_{\rm max}$ for
positive defects mirrors the symmetry of the obstacles but drops nearly to zero partway across the region connecting the left and right sides of the obstacles, indicating that $+1/2$ defects are no longer flowing horizontally. Meanwhile, the $-1/2$ defects become even more strongly localized in the region between obstacles.

We find that at large gap sizes, the motion of the defects tends to mirror the flow of the system, but that the defects may become pinned for a period of time, reducing their flow speed. As the gap size decreases, the likelihood of pinning diminishes and the defects can move more freely. At a critical gap size of $d = 5$, the negative defects become strongly localized in the vertical gap between obstacles. At first this allows the positive defects to travel more efficiently in the $-x$ direction by skirting the negative defects, but as the gap size diminishes further, the positive defects begin to annihilate with the negative defects and the $x$-direction flow is lost. Instead, the $+1/2$ defects begin to travel transverse to the flow in lanes along the $y$-direction, forming a smectic-like defect state. In Supplemental Movies 1--3 we show the nematic configuration and flow velocities for simulations in these three regimes.

Since we find maximal $-x$ direction flow at $d = 5$ for all values of $\alpha$, the maximum does not result from a commensuration effect between $d$ and the active length scale $\xi_a \propto 1/\sqrt{\alpha}$.
Instead, the commensuration occurs between $d$ and
the characteristic size of topological defects $\xi_d$, which we hold fixed in this study.
It appears when $d$ reaches a length for which
$-1/2$ defects become localized between the obstacles,
enhancing the overall flow.

{\it Conclusion---} We showed numerically that a periodic array of asymmetric obstacles can produce translational ratchet flows in an active nematic. As the gap distance between obstacles decreases, the flow velocity directions become peaked along the asymmetry direction, but the average flow velocity varies non-monotonically. The obstacle asymmetry induces a translational active nematic ratchet that has not been described previously. The ratcheting effect is robust over a large range of obstacle gap sizes and activity levels. We also observed a transition in defect dynamics that is correlated with the flow speed non-monotonicity. Positive winding defects follow the fluid flow for large gap sizes, while for small gap sizes, pinned negative defects inhibit the movement of positive defects along the flow and cause the positive defects to travel transverse to the flow.

This work opens a variety of future directions for steering or patterning active nematic flows and defect structures using ratchet geometries. Such effects have potential microfluidic applications, including logic gate design \cite{woodhouse17} or the creation of complex patterns \cite{Jorge24}. It would be interesting to explore other asymmetric obstacle geometries or lattice arrangements. Different obstacle geometries may produce distinct topological defect arrangements, while different lattices may generate novel flow patterns.

\begin{acknowledgements}
This work was supported by the U.S. Department of Energy through the Los Alamos National Laboratory. Los Alamos National Laboratory is operated by Triad National Security, LLC, for the National Nuclear Security Administration of the U.S. Department of Energy (Contract No. 89233218CNA000001).
\end{acknowledgements}

\bibliography{LC}

\end{document}


\title{Supplemental Material for ``Active Nematic Ratchet in Asymmetric Obstacle Arrays''}

\author{Cody D. Schimming}
\email[]{cschim@lanl.gov}
\affiliation{Theoretical Division and Center for Nonlinear Studies, Los Alamos National Laboratory, Los Alamos, New Mexico, 87545, USA}

\author{C. J. O. Reichhardt}
\affiliation{Theoretical Division and Center for Nonlinear Studies, Los Alamos National Laboratory, Los Alamos, New Mexico, 87545, USA}

\author{C. Reichhardt}
\affiliation{Theoretical Division and Center for Nonlinear Studies, Los Alamos National Laboratory, Los Alamos, New Mexico, 87545, USA}

\maketitle

\section{Active nematic model and numerical method}

For details on the continuum model and numerical method, see the Supplemental Material for Ref. \cite{schimming24}.

\section{Defect Statistics}

To measure the number of defects in the system at a given time we integrate the magnitude of the two-dimensional topological defect density $|D|$ \cite{blow14,schimming22}:
\begin{equation}
    N = \int | \varepsilon_{k \ell} \varepsilon_{\mu \nu} \partial_k Q_{\mu \alpha} \partial_{\ell} Q_{\nu \alpha} | \, d\mathbf{r}
\end{equation}
where $\bm{\varepsilon}$ is the two-dimensional Levi-Civita tensor and summation on repeated indices is assumed. The averages shown in Fig. S1 are averaged over the simulation after it has reached a dynamical steady state. To get the positional distribution of defects shown in Figs. 5 and S2, we first find locations with nematic scalar order parameter $S < 0.1 S_N$ and then find the locations of maximal $|D|$ within that subset. To determine whether a defect is $+1/2$ or $-1/2$ winding number, we measure $\text{sign}(D)$ at the defect location. For each type of defect, we first collect the number of defects at each point relative to each obstacle, then sum over all obstacles. The data presented in Fig. 5 is normalized by the maximum count, $N_{max}$, after summing over all obstacles. 

To get the distribution of defect velocity directions shown in Fig. 4, we measure the topological current $\mathbf{J}$ \cite{schimming22,schimming23} at the location of defects:
\begin{equation}
    J_i = \varepsilon_{i k} \varepsilon_{\mu\nu} \partial_t Q_{\mu \alpha} \partial_k Q_{\nu \alpha}.
\end{equation}
To ensure we only count defects that are moving, the statistics in Fig. 4 only include defects for which $|\mathbf{J}| > 0.1|\mathbf{J}|_{max}$.

\section{Supplemental Movies}
\begin{itemize}
    \item Supplemental Movie 1: (left) Nematic configuration and (right) velocity and vorticity for a simulation with $d = 8$ and $\alpha = 1.5$. The color in the left movie represent the scalar order parameter $S$ and the white lines represent the director $\mathbf{n}$. The color in the right movie represents the vorticity $\omega$ while the black arrows represent the velocity field $\mathbf{v}$.

    \item Supplemental Movie 2: (left) Nematic configuration and (right) velocity and vorticity for a simulation with $d = 5$ and $\alpha = 1.5$. The color in the left movie represent the scalar order parameter $S$ and the white lines represent the director $\mathbf{n}$. The color in the right movie represents the vorticity $\omega$ while the black arrows represent the velocity field $\mathbf{v}$.

    \item Supplemental Movie 3: (left) Nematic configuration and (right) velocity and vorticity for a simulation with $d = 2$ and $\alpha = 1.5$. The color in the left movie represent the scalar order parameter $S$ and the white lines represent the director $\mathbf{n}$. The color in the right movie represents the vorticity $\omega$ while the black arrows represent the velocity field $\mathbf{v}$.
    
\end{itemize}

\newpage
\section{Supplemental Figures}

\renewcommand{\thefigure}{S1}
\begin{figure}[h]
    \centering
    \includegraphics[width = \textwidth]{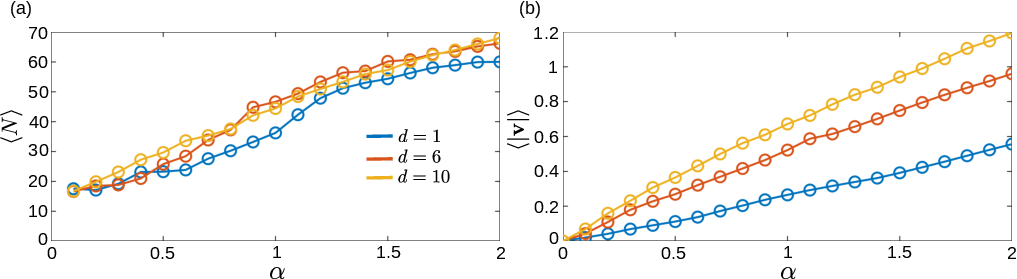}
    \caption{(a) Average number of topological defects $\langle N \rangle$ versus activity $\alpha$ for simulations with $d = 1$, $d = 6$, and $d = 10$. (b) Average flow speed $\langle |\mathbf{v}|\rangle$} versus $\alpha$ for simulations with $d = 1$, $d = 6$, and $d = 10$. Both quantities scale roughly linearly with activity, indicating systems in active turbulence.
    \label{fig:DefectNumber}
\end{figure}

\renewcommand{\thefigure}{S2}
\begin{figure}[h]
    \centering
    \includegraphics[width = \textwidth]{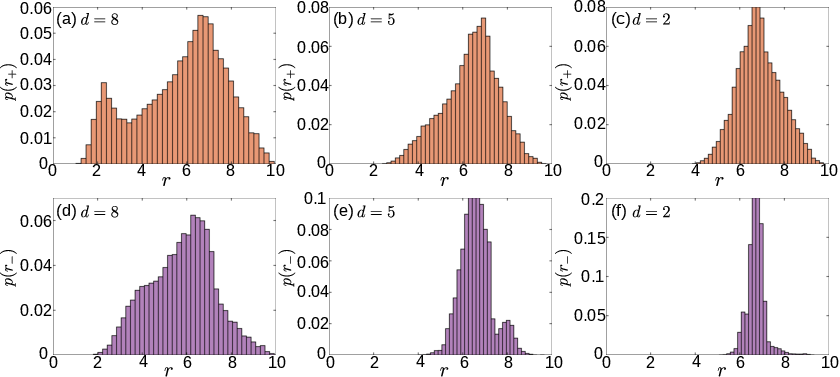}
    \caption{Distribution of radial distance of defects from the center of an obstacle for (a--c) $+1/2$ defects and (d--f) $-1/2$ defects. Shown for simulations with (a,d) $d = 8$, (b,e) $d = 5$, and (c,f) $d = 2$.}
    \label{fig:DefectRadialDist}
\end{figure}

\newpage

\bibliography{LC}